\documentclass[showpacs,aps,prb,twocolumn,longbibliography,superscriptaddress,10pt]{revtex4-1}
\usepackage{graphicx} % Include figure files
\usepackage{bm}% bold math
\usepackage{color}
\usepackage{amsmath}
\usepackage{amssymb}
\usepackage{enumerate}
\usepackage{xspace}
\usepackage{mathrsfs}
\usepackage{mathptmx}
\usepackage[colorlinks=true,linkcolor=blue,citecolor=blue,urlcolor=blue]{hyperref}
\usepackage[utf8]{inputenc}

\newcommand{\kN}{\ensuremath{k_{\mathrm{N,i}}}\xspace}

\newcommand{\smJij}{\ensuremath{J_{ij}}\xspace}

\newcommand{\smB}{\ensuremath{\mathbf{B}}\xspace}

\newcommand{\sms}{\ensuremath{\mathbf{S}}\xspace}
\newcommand{\e}{\ensuremath{\mathbf{e}}\xspace}
\newcommand{\vampire}{\textsc{vampire} }

\begin{document}

\title{Local setting of spin textures in a granular antiferromagnet}
\author{Miina~Leiviskä}%\orcidlink{0000-0002-3815-2267}}
\email{miina.leiviska@cea.fr}
\affiliation{Univ. Grenoble Alpes, CNRS, CEA, Grenoble INP, IRIG-Spintec, F-38000 Grenoble, France}
\author{Sarah~Jenkins}%\orcidlink{0000-0002-6469-9928}}
\affiliation{School of Physics, Engineering and Technology, University of York, York YO10 5DD, United Kingdom}
\author{Richard~F.~L.~Evans}%\orcidlink{0000-0002-2378-8203}}
\email{richard.evans@york.ac.uk}
\affiliation{School of Physics, Engineering and Technology, University of York, York YO10 5DD, United Kingdom}
\author{Daria~Gusakova}%\orcidlink{0000-0002-5111-7079}}
\affiliation{Univ. Grenoble Alpes, CNRS, CEA, Grenoble INP, IRIG-Spintec, F-38000 Grenoble, France}
\author{Vincent~Baltz}%\orcidlink{0000-0002-7637-1938}}
\email{vincent.baltz@cea.fr}
\affiliation{Univ. Grenoble Alpes, CNRS, CEA, Grenoble INP, IRIG-Spintec, F-38000 Grenoble, France}
\begin{abstract}
Controlling the magnetic order of antiferromagnets is challenging due to their vanishing net magnetization. For this reason, the study of local spin textures in antiferromagnets is restricted by the difficulty in nucleating such states. Here, using atomistic simulations we demonstrate a method for nucleating localized spin textures in the grains of thin film antiferromagnet, $\gamma$-IrMn$_3$. Utilising the exchange bias coupling between a ferromagnet and an antiferromagnet, we set the spin texture in the latter from a predefined spin texture in the former by means of a thermal cycling procedure. The local textures set in the antiferromagnetic grains are shown to be stable against field perturbations. We also discuss how various material parameters affect the efficiency of the setting and the characteristics of these set textures. The setting of antiferromagnetic spin textures provides a potential route to antiferromagnetic spintronic devices with non-collinear spin states such as skyrmions, bubbles and domain walls. 
\end{abstract}

%\pacs{75.10.Hk,75.20.-g,75.50.Ss,75.60.Jk,75.78.Jp}
\maketitle

\section{Introduction}
The interest in antiferromagnetic (AFM) states is founded on their robustness against perturbations, such as magnetic fields, as well as on their physical properties that extend the potential of spintronics beyond that based on ferromagnets (FM) only \cite{Jungwirth2016,Baltz2018}. Analogously to FMs, local spin textures with inhomogeneous magnetization configurations exist in AFMs under conditions yielding a favorable energy balance. Due to the different local symmetries, the expected properties of these AFM spin textures differ from and have interesting advantages over their ferromagnetic counterparts \cite{Gomonay2018}. However, due to the lack of net magnetization, the main barrier to accessing these properties is the difficulty in experimentally nucleating them. Typically, structures with a size comparable to the critical single domain size or domain wall width are patterned to make the global multidomain structure of continuous films energetically unfavorable so that local spin textures form instead \cite{Salazar-Alvarez2009,Wu2011}. Alternative methods for the nucleation of local AFM spin textures have also been proposed: through injection of vertical spin-polarized current in nanostructured devices \cite{Zhang2016afm} or through subjecting the system to ultrafast laser pulses \cite{Khoshlahni2019}.

\begin{figure}[!hb]
\includegraphics[width=0.45\textwidth, trim=0 0 0 0]{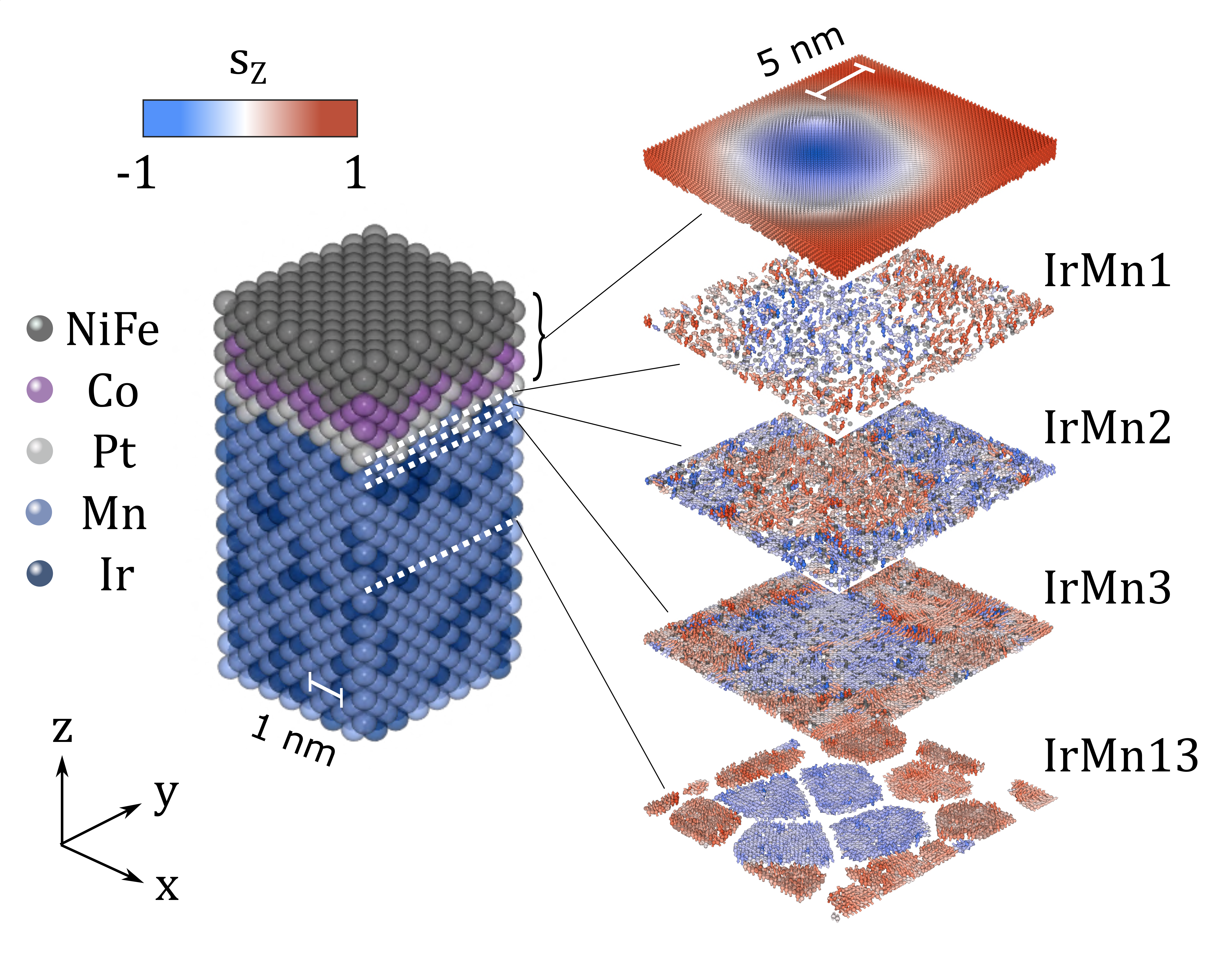}
\caption{(Color Online) (Left) Atomic structure of a slice of the simulation system consisting of a hybrid NiFe/Co/Pt FM single crystal coupled to a granular $\gamma$-IrMn$_3$ AFM across an atomically intermixed interface. The full size of the simulation system is $100 \times 100 \times 6.67$ nm. (Right) Cross-section of the spin texture obtained in the FM and at several depths in the AFM, following a thermal cycling procedure. The numbers indicate the relative position of the AFM monolayers away from the interface.}
\label{fig:stack}
\end{figure}

Here, we use atomistic simulations to demonstrate a local nucleation of spin textures in AFM grains that does not require complex device geometries and is applicable beyond metallic AFMs. Utilizing the exchange bias coupling at a FM/AFM interface and a thermal cycling procedure, we set the spin texture in the AFM grains during their magnetic ordering from predefined spin textures in the FM. This method follows previous experimental works on setting AFM domains and domain walls \cite{VallejoFernandez2010}, vortices \cite{Salazar-Alvarez2009,Wu2011}, and bubbles \cite{Rana2021}. In this work, we nucleate prototypical spin textures, namely skyrmion bubbles (Sk), in a continuous thin film FM using a carefully optimized magnetic stack and an external field. We then use this as a template for the setting of the underlying AFM grains, as shown in Figure \ref{fig:stack}. Note, however, that the spin textures set in the AFM grains are not topologically protected despite the initial template comprising of Sk bubbles.

\section{The simulation system}
The simulation system, Ni$_{80}$Fe$_{20}$ (0.87)/Co (0.33)/Pt (0.47)/$\gamma$-IrMn$_3$ (5 nm), is based on an experimental one that has been optimized for the nucleation of Sks in the FM layer \cite{Rana2021,Boulle2016}. The ultrathin Co/Pt bilayer provides a strong interfacial Dzyaloshinskii-Moriya interaction (DMI) \cite{Yang2015} and some perpendicular magnetic surface anisotropy, while the thickness of the NiFe layer allows the vanishing of the total anisotropy. The magnetic coupling between the FM and AFM layers is preserved due to finite atomic intermixing. For the AFM layer we use the non-collinear $\gamma$-IrMn$_3$, due to its large exchange bias coupling originating from the statistical imbalance of spins in the four magnetic sublattices, which yields uncompensated interfacial spins~\cite{Jenkins2020}. Moreover, the four easy axes~\cite{Jenkins2019}
of IrMn$_3$ may facilitate the magnetic moment winding that is necessary for a non-collinear spin texture formation. The magnetic properties of $\gamma$-IrMn$_3$ relating to its non-trivial spin structure have been thoroughly characterized using the atomistic simulation software package \vampire \cite{Jenkins2021,Evans2014}, which is also used for this work. The simulations utilized the ARCHER2 supercomputer with typical simulations running on 1024 CPU cores due to the large number of Monte Carlo time steps and large system size of around $6 \times 10^6$ spins.

The spin Hamiltonian governing the energetics of the system is given by
\begin{equation}
\begin{split}
\mathscr{H} = -\sum_{i<j} \smJij \sms_i \cdot \sms_j &- \sum_{i< j}^z \kN (\sms_i \cdot \e_{ij})^2  - \sum_{i<j}\mathbf{D}_{ij}\cdot(\sms_i \times \sms_j) \\
&- \sum_i \mu_{s,i} \sms_i \cdot \smB_{ext}.
\label{eq:hamiltonian}
\end{split}
\end{equation}
The first term is the exchange interaction, where subscripts \textit{i} and \textit{j} refer to spins on sites $i$ and $j$, $J_{ij}$ is the exchange constant and $\sms_i$ is the spin vector. The second term describes the N\'eel pair anisotropy that rotates the spins radially/tangentially with respect to a nearest neighbor of a given element and is used to simulate the interfacial perpendicular magnetic anisotropy (Co spins point radially to Pt sites) and correct anisotropy of IrMn$_3$ (Mn spins point tangentially to Ir sites) \cite{Jenkins2019}. Here, \kN is the N\'eel pair anisotropy constant, $\e_{ij}$ is a unit vector between atoms $i$ and $j$, and $z$ is the number of nearest neighbors. The third term that stabilizes the topological states is the DMI interaction where $\mathbf{D}_{ij}$ is the DMI vector, defined as $\mathbf{D}_{ij}=D_{ij}(\mathbf{\hat{r}}_{ik}\times\mathbf{\hat{r}}_{jk})$, where $D_{ij}$ is the DMI strength and $\textbf{\^{r}}_{ik,jk}$ is the unit vector between the magnetic atoms at site $i$ and $j$ and Pt at site $k$.
\begin{figure}[!hb]
\includegraphics[width=0.49\textwidth, trim=0 0 0 0]{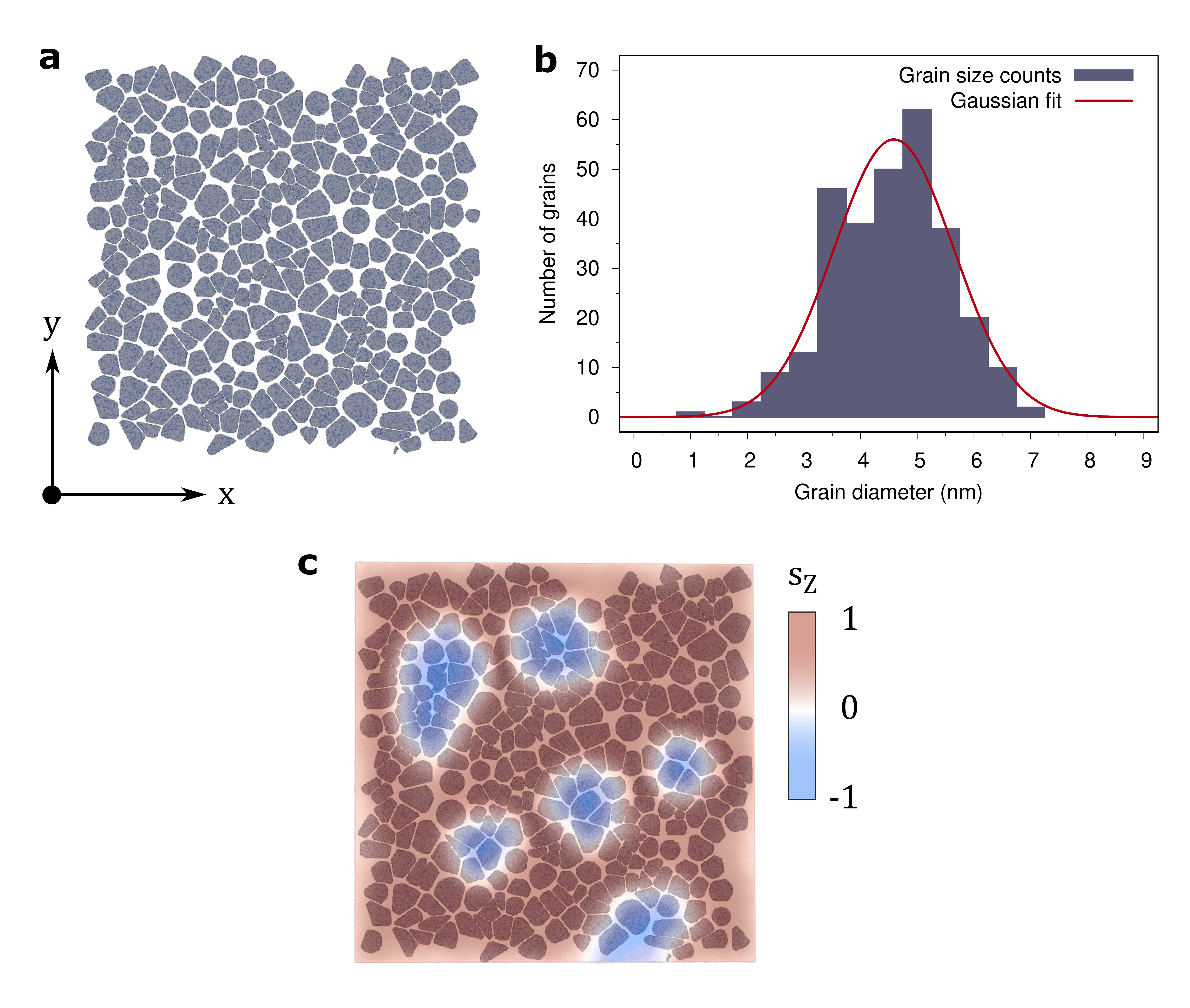}
\caption{(Color Online) (a) The bare grain structure of the AFM layer created using a Voronoi construction. (b) The grain size distribution showing an average grain size of $\sim$ 4.5 nm. (c) Overlaying the FM layer spin structure and the bare grain structure shows that the irregular shape of the skyrmions in the former follows the grain boundaries of the latter.}
\label{fig:grains}
\end{figure}
The fourth term is the Zeeman energy where $\mu_{s,i}$ is the atomic magnetic moment and $\smB_{ext}$ is the external induction or flux density. For the sake of computational simplicity, we have not included the RKKY-interaction, which would increase the exchange interaction strength between the FM and AFM layers but not change the physics of the local setting presented in this work. 

The temperature-dependence of the magnetic state is simulated using an adaptive Monte Carlo Metropolis algorithm~\cite{AlzateCardona2019}, which is ideal for naturally simulating the temperature-evolution of the magnetic state. The field-dependence, on the other hand, was simulated using the stochastic Landau-Lifshitz-Gilbert equation and a Heun numerical scheme \cite{GarciaPalacios1998}. We have set the damping parameter to 1 to promote rapid relaxation and therefore faster numerical calculation, since our goal is to reach the ground state rather than follow the exact dynamics. The timestep is 0.1 fs.

The simulation system with a size of 100x100x6.67 nm$^3$ has a layer-dependent structure: the FM/Pt multilayer is a single crystal while the AFM layer is granular with an average grain size and spacing of 5 and 0.75 nm, respectively, in line with experimental data \cite{VallejoFernandez2008}. Similarly to an experimental system, the AFM grains are not coupled to each other \cite{OGrady2010}. On the other hand, the FM grains are coupled and as the FM and AFM grains are coincident, the FM layer can be modeled as a single crystal. The grain structure (illustrated in Figure \ref{fig:grains}) is created using a Voronoi construction as described previously in Ref. \cite{Jenkins2021b}. It should be noted that we have not implemented periodic boundary conditions as they are incompatible with the implementation of the AFM grain structure. The crystal orientation of the entire simulation stack is such that the [001] crystal axis is along the film normal.

The onsite magnetic properties of each magnetic element in the simulation stack are listed in Table \ref{tab:parameters}. For the permalloy layer (Ni$_{80}$Fe$_{20}$), $\mu_s$ was calculated from the saturation magnetization $M_s$ of $\sim 8.6\times 10^{5}$ A/m using $\mu_s = M_s a^3/n_{at}$ \cite{Evans2014}, where $a$ is the lattice constant and $n_{at}$ is the
\begin{table}[!hb]
\centering % used for centering table
\begin{tabular}{l c c c c} % centered columns (4 columns)
Parameter & Ni$_{80}$Fe$_{20}$ & Co & Mn\\
\hline\hline % inserts single horizontal line
$\mu_s$ ($\mu_B$)            & 1.05      & 1.72          & 2.6 \\
\hline
$k_N$  (10$^{-22}$ J/atom)   & $-$       & 0.025 (Pt)    & -4.22 (Ir) \\
\hline
Intermixing (\AA)            & 1.67      & 1.67          & 3.34 \\
\hline
\end{tabular}
\caption{Parameters used in the simulation for each magnetic element.}
\label{tab:parameters}
\end{table}
\begin{table}[!hb]
\centering % used for centering table
\begin{tabular}{c|c c c c} % centered columns (4 columns)
 & NiFe & Co & Pt & Mn\\
\hline % inserts single horizontal line
NiFe    & 3.8 \cite{Evans2014}  & 5    & 0.5*  & 1.21 \cite{Szunyogh2009} \\
Co      & 5    & 6.1 \cite{Evans2014}  & 0.5*  & 1.21 \cite{Szunyogh2009} \\
Pt      & 0.5* & 0.5* & $-$   & $-$  \\
Mn      & 1.21 \cite{Szunyogh2009} & 1.21 \cite{Szunyogh2009} & $-$   &\begin{tabular}{c} -6.4 (NN) \cite{Szunyogh2009} \\ 5.1 (NNN) \cite{Szunyogh2009} \end{tabular} \\
\end{tabular}
\caption{Matrix of exchange (\smJij) and DMI* coefficients ($D_{ij}$) (10$^{-21}$ J/link) for the interactions between the different elements in the stack. NN stands for nearest neighbors and NNN for next nearest neighbors}
\label{tab:exchangecoeff}
\end{table}
number of atoms in the unit cell. For Co the values were obtained from Ref. \cite{Evans2014} and for $\gamma$-IrMn$_3$ from Ref. \cite{Jenkins2020}. The intermixing in Tables \ref{tab:parameters} refers to the thickness over which the given layer is intermixed with the one below. The intersite exchange ($J_{ij}$) and DMI ($D_{ij}$) interactions are listed in Table \ref{tab:exchangecoeff}. For Ni$_{80}$Fe$_{20}$ the $J_{ij}$ is calculated from the expected Curie temperature $T_c$ of $\sim 900$ K using $J_{ij}=3k_BT_c/\epsilon z$, where $\epsilon=0.79$ is the spin-wave mean-field correction and $z$ is the number of nearest neighbors \cite{Evans2014}. For Co the values are obtained from Ref. \cite{Evans2014} and for $\gamma$-IrMn$_3$ from \textit{ab initio} calculations \cite{Szunyogh2009}. For the coupling strength between Co and NiFe, we have used the average Co-Co and NiFe-NiFe coupling strengths. The DMI value used in our calculations is $0.5\times 10^{-21}$ J/link, which corresponds to an interfacial DMI parameter $D_s$ = 0.5 pJ/m for the limiting case of a flat interface.  Experimental values available in the literature for similar stacks (measured by Brillouin light scattering) range from $\sim$ 0.15 to $\sim$ 1.25 pJ/m \cite{Ma2017, Khan2018, Rana2021}. We have verified that varying the DMI value within this range does not alter the main conclusions of the paper.

\section{Local setting of the AFM grains}

\begin{figure}[!hb]
\includegraphics[width=0.49\textwidth, trim=0 0 0 0]{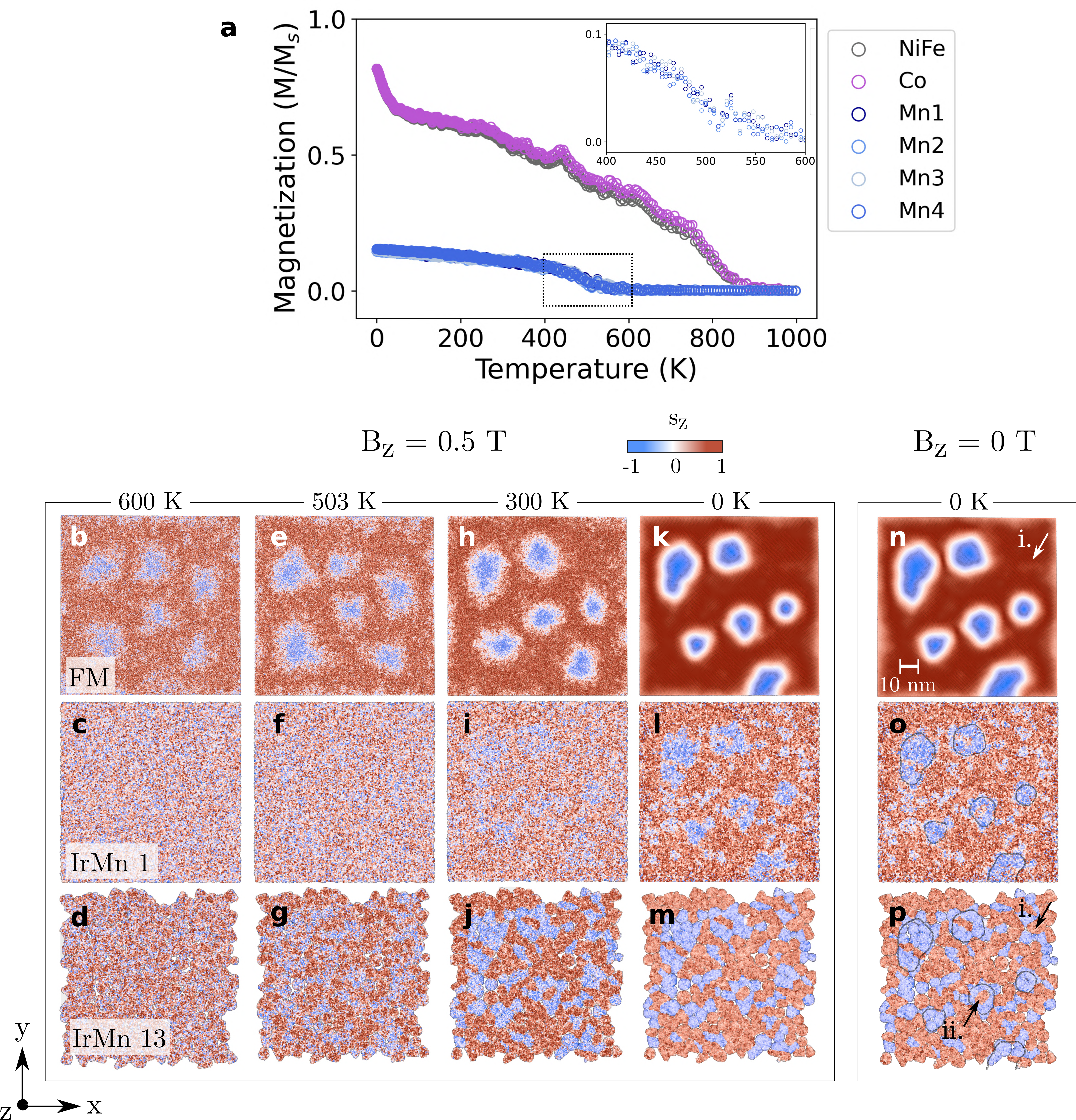}
\caption{(Color Online) (a) Temperature dependence of the magnetization (M/Ms) of each magnetic material in the simulation stack. (b-m) Top view snapshots of the evolution of the spin textures in the FM (Top row), AFM at the interface (Middle row, IrMn 1) and AFM in the core (Bottom row, IrMn 13), under out-of-plane field-cooling (B$_z$=0.5 T): at 600 K, above the Néel temperature T$_N$ of the AFM; at 503 K, near T$_N$; at room temperature; and 0 K well below T$_N$. (n-p) The spin textures at remanence, for B$_z$=0 T. The contours of the FM textures in n) are superimposed to o) and p), as visual guides.} 
\label{fig:imprinting}
\end{figure}

The localized setting of the AFM grains is realized by means of thermal cycling. More precisely, the simulations were carried out in three steps: i) FM Sk nucleation, ii) field-cooling, and iii) field removal. During the first step (Figure~\ref{fig:imprinting}b-d), the system was equilibrated at 600 K in an applied field of B$_z$ = 0.5 T parallel to the film normal for $10^6$ Monte-Carlo steps. Because 600 K is above the N\'eel temperature (T$_N$) of the AFM layer ($\sim$ 510 K, see Figure \ref{fig:imprinting}a) and below the Curie temperature of the FM layer ($\sim 900 $ K, see Figure \ref{fig:imprinting}a), this phase allows the nucleation of nanoscale Sks in the FM layer while the AFM layer remains disordered throughout. This ensures that the setting of spin textures is from the FM to the AFM layer and not vice versa.

\begin{figure}[!ht]
\includegraphics[width=0.49\textwidth, trim=0 0 0 0]{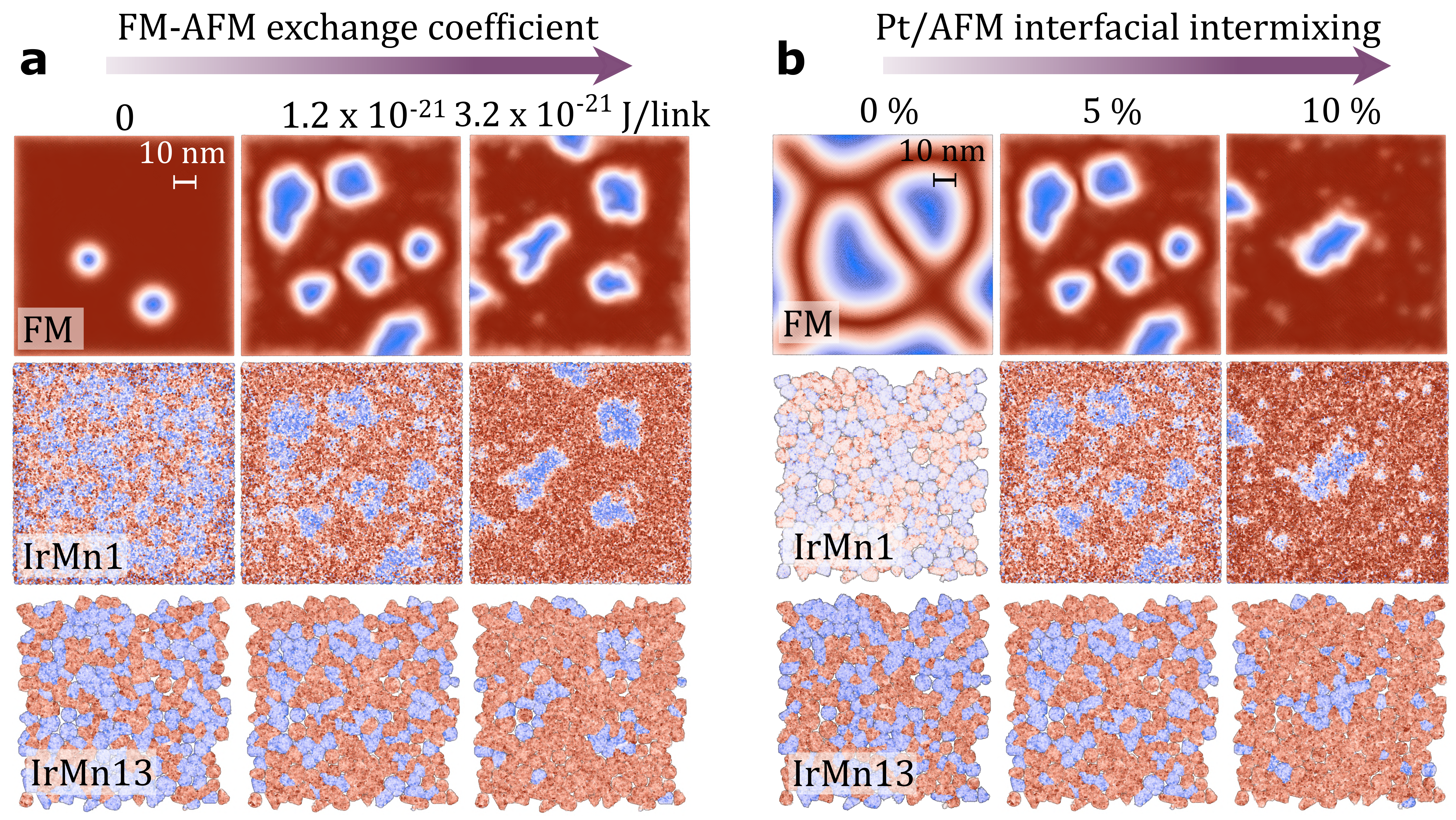}
\caption{(Color Online) The effect of increasing (a) the exchange coefficient strength and (b) the intermixing between the Pt and AFM layers on the imprinting efficiency in terms of FM-AFM spin texture conformity and absence of discrepancies.}
\label{fig:excinterm}
\end{figure}
In the second simulation step (Figure \ref{fig:imprinting}e-g), the system is cooled down to 0 K in the same field in order to further stabilize the FM layer and to set the AFM layer upon crossing the N\'eel temperature $T_N$. The bulk AFM (exemplified by IrMn layer 13 in Figure \ref{fig:imprinting}) begins to set at the blocking temperature of the IrMn grains around $T \approx 500$ K. The interfacial moments of the AFM (IrMn 1 in Figure \ref{fig:imprinting}), on the other hand, are subject to more thermal noise and the setting of the AFM spin textures is more difficult to discern. This is likely due to the reversible component fluctuating with the FM spins. At $T = 0$ K (Figure \ref{fig:imprinting}k-m), the setting of the  the AFM interface and bulk based on the FM spin textures becomes clear: the textures observed in the FM are largely reproduced in the bulk AFM as highlighted by the contours in Figure \ref{fig:imprinting}o,p and their exact shape is governed by the grain boundaries of the AFM.

Superimposed to the set spin textures there are also random spin textures in the AFM bulk that don't correspond to the FM spin configuration. These discrepancies can be divided into two categories based on their location and the extent of deviation: discrepancies exemplified by i. in Figure \ref{fig:imprinting}n,p are in the region of uniformly magnetized FM and often accompanied by slight impressions in the adjacent FM layer. Discrepancies exemplified by ii. in Figure \ref{fig:imprinting}p are inside the imprinted spin textures and deviate completely from the FM spin configuration. The reason for these discrepancies is likely due to the FM-AFM exchange interaction not being strong enough to set all the grains, which is demonstrated in Figure \ref{fig:excinterm}a - stronger exchange interaction improves the FM-AFM spin texture conformity while zero exchange interaction results in the AFM having a completely random domain structure. A similar outcome is realized when the intermixing at the Pt/AFM interface is increased, as shown in Figure \ref{fig:excinterm}b because more FM and AFM spins will come to a direct contact, which facilitates the AFM-FM exchange interaction. We recall that for the sake of computational simplicity we have not included the RKKY-interaction. This would increase the exchange interaction strength between the FM and AFM layers and give rise to non-zero exchange bias in the absence of intermixing but not change the physics of the local setting presented in this work. It is also possible that some of the i. type discrepancies are residual imprintings of FM Sks that moved or were annihilated during the field-cooling (compare the FM Sk locations in Figures \ref{fig:imprinting}b and \ref{fig:imprinting}n). We also point out that layer intermixing is likely to finely modify some magnetic properties \cite{Baltz2013} and parameters like DMI, which can also alter the imprinting efficiency.

\begin{figure*}[!ht]
\includegraphics[width=0.8\textwidth]{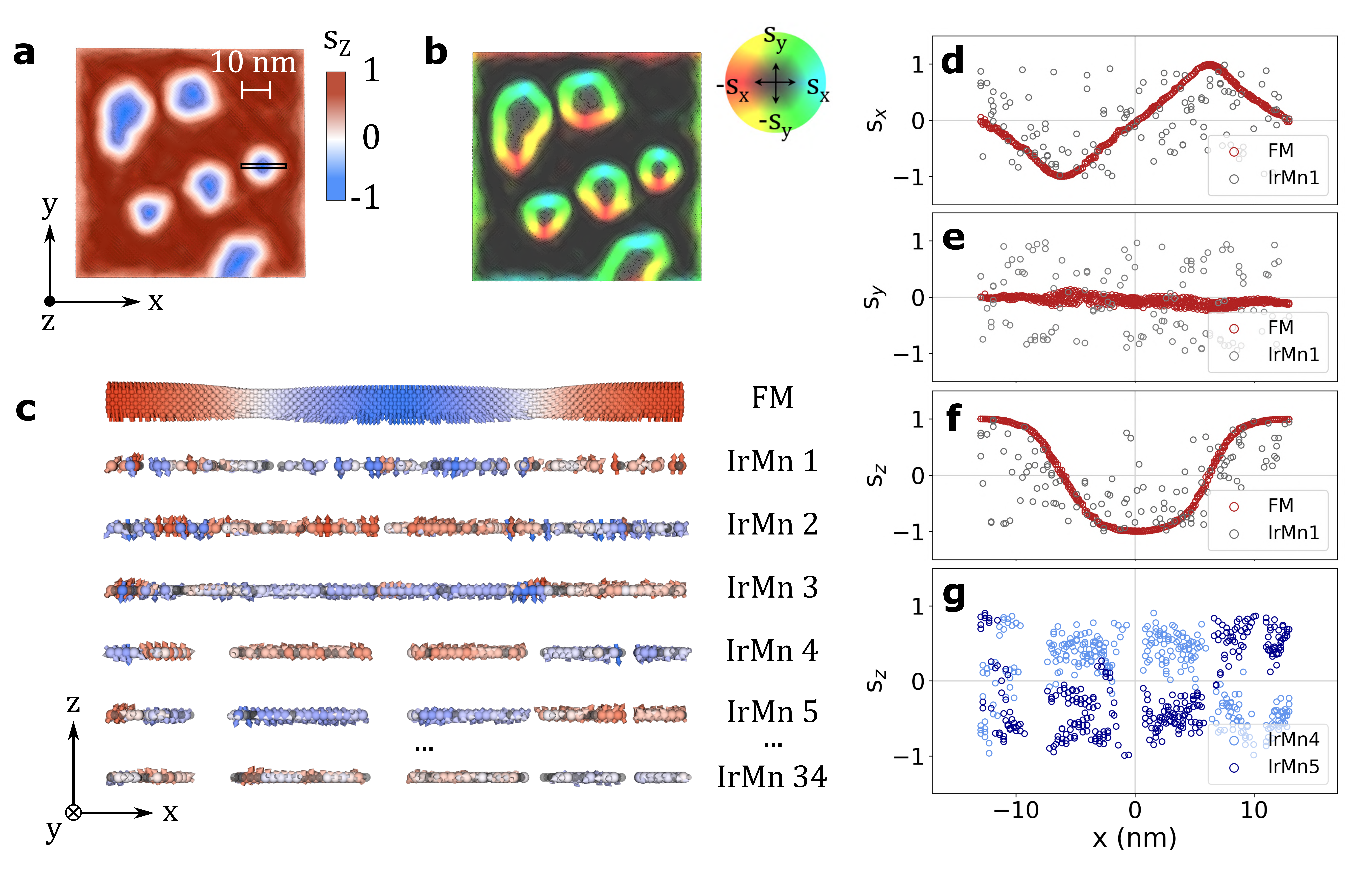}
\caption{(Color Online) Top view of the spin structure in the FM layer showing azimuthal (a) and rotational (b) components of the magnetization showing the existence of Sk bubbles in the ferromagnetic layer. (c) Layer-by-layer spin structure through the sample, corresponding to the cross-section indicated in (a) showing the propagation of the spin texture into the antiferromagnet. The spin direction alternates in the alternating AFM monolayers due to the antiferromagnetic coupling. (d-g) Lateral dependencies of the spin components $s_i$ in the FM, IrMn 1, IrMn 4 and IrMn 5 layers showing the presence of domain walls in the FM and corresponding structure in the antiferromagnet.}
\label{fig:spintext}
\end{figure*}

Finally, the last simulation step shown in Figure \ref{fig:imprinting}n-p is the removal of the external field, which causes only minor changes to the spin textures throughout the stack, demonstrating the zero-field stability of the set localized AFM spin textures. 

In summary, this set of simulations shows that i) spin textures (here Sks) initially nucleated in a FM layer serve as a template for the localized setting of spin textures in the AFM grains during a thermal cycling protocol, ii) the setting of the AFM textures penetrates into the bulk of the AFM and iii) the AFM textures are stable at remanence. Note that these conclusions are valid for any final temperature below $T_\mathrm{N}$, including room temperature in our case (see Figure \ref{fig:imprinting}h-j). The robustness and universality of our results is later further demonstrated in Figure \ref{fig:maze}b where we show the setting of the AFM grains based on another FM spin texture, namely maze domains.

\section{Microscopic details of the local setting}

The exact nature and morphology of  the set textures (Figure \ref{fig:spintext}) is discussed next. The radii of the Sks in the FM (Figure \ref{fig:spintext}a-b) range from $\sim$ 10 to 20 nm and their shape is rather irregular, most likely due to the pinning by the grain boundaries of the AFM layer (see Figure \ref{fig:grains}c for the conformity between the Sk shapes and the AFM grain structure). The variation of the spin components $s_x$, $s_y$, and $s_z$ across the cross-section shows behavior characteristic of a N\'eel type Sk (Figure \ref{fig:spintext}d-f), which is in good agreement with the interfacial DMI, and its strength being above the critical strength for a Bloch  to  N\'eel  domain  wall  transition ($D_c=4\mu_0M_s^2t\ln{2}/2\pi^2$). The $s_x$ and $s_z$ components can be fitted using $\cosh^{-1}(\pi[x-x_0]/\Delta)$ and $\tanh(\pi[x-x_0]/\Delta)$ functions, respectively, where $x_0$ is the domain wall center location and $\Delta$ is the domain wall width. For the selected Skyrmion, $\Delta \sim 8.4 $ nm, which is in the same order of magnitude as expected by theory \cite{Wang2018}. 

\begin{figure}[!hb]
\includegraphics[width=0.49\textwidth, trim=0 0 0 0]{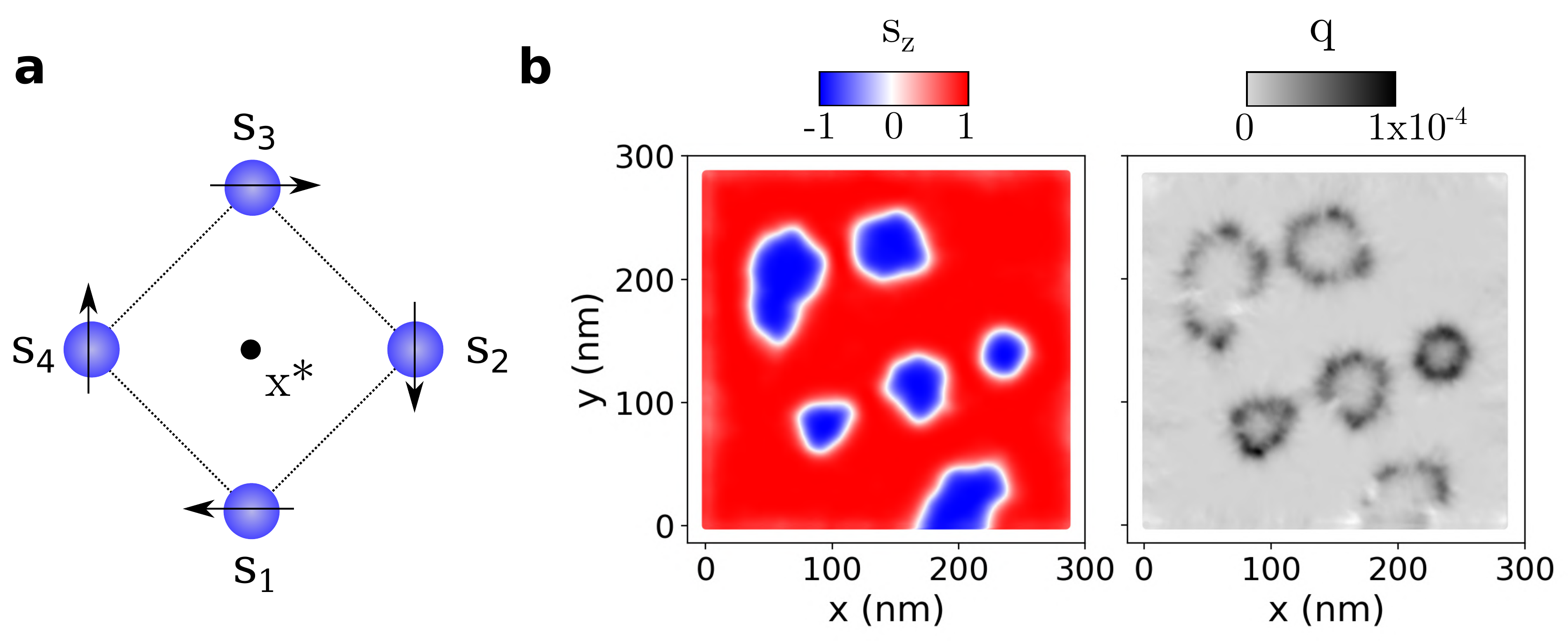}
\caption{(Color Online) a) Schematic showing the method for calculating the topological charge density $q$ of a discrete lattice made up of square unit cells. The total topological charge $Q$ is the sum of $q$ over the entire lattice. b) The topological charge density calculated for the FM layer.}
\label{fig:winding}
\end{figure}

We calculate the winding number $Q$ for our discretized system by summing up the topological charge density $q$ per a square cell illustrated in Figure \ref{fig:winding}a:
\begin{equation}
    q(x^*) = \frac{1}{4\pi}[(\sigma A)(s_1,s_2,s_3)+(\sigma A)(s_1,s_3,s_4)],
\end{equation}
\begin{equation}
    \sigma A(s_j,s_k,s_l) = 2\tan^{-1}\left(\frac{\mathbf{s}_j\cdot\mathbf{s}_k\times\mathbf{s}_l}{1+\mathbf{s}_j\cdot\mathbf{s}_k+\mathbf{s}_j\cdot\mathbf{s}_l+\mathbf{s}_k\cdot\mathbf{s}_l}\right),
\end{equation}
where $\sigma A$ is the signed area of a triangle delimited by $\mathbf{s}_j$, $\mathbf{s}_k$, and $\mathbf{s}_l$, $\mathbf{s}_i$ being the magnetization vector at site $i$ \cite{Berg1981,Bottcher2018}. Figure \ref{fig:winding} shows the calculation of the winding number for the FM layer in Figure \ref{fig:imprinting}n, which yields $Q = 1$ for each Sk bubble, confirming their topological nature. It should be noted that the FM spin texture at the boundary of the simulation system, is not a Sk bubble (non-integer Q $\sim$ 0.4) and thus not topologically protected. Considering that the sum of the topological charge over the whole lattice should return an integer, the winding number of 0.4 of the FM texture should be compensated elsewhere. As the topological charge is only being calculated for the FM layer, it is possible that the AFM spins compensate the fractional Q or, alternatively, the existence of the boundary and pinning from the AFM may cause the fractional Q as technically Q is only defined for infinite systems.

\begin{figure}[!ht]
    \centering
    \includegraphics[width=0.45\textwidth, trim=0 0 0 0]{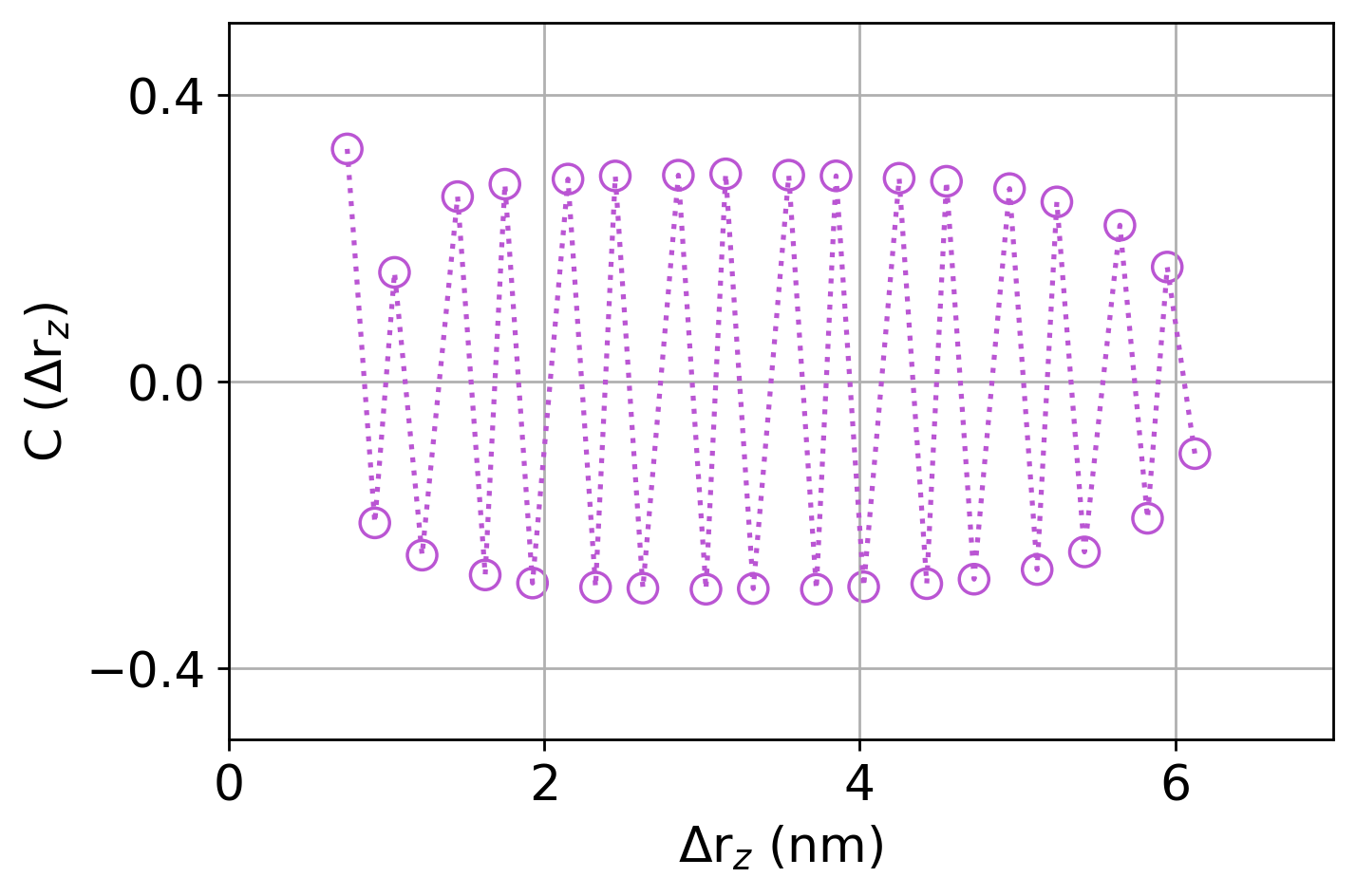}
    \caption{(Color Online) Partial spin-spin correlation function of the spin texture showing alternating correlation in the $s_z$ spin components between the ferromagnet and the antiferromagnet.}
    \label{fig:ssc}
\end{figure}

At the AFM interface (IrMn 1 in Figure \ref{fig:spintext}), the $s_x$, $s_y$, and $s_z$ components show average behavior that corresponds to that of the FM layer, providing further support for the AFM spin texture setting from the FM template. The increased noise is likely due to the strong anisotropy of $\gamma$-IrMn$_3$, interfacial intermixing, and the interfacial magnetization comprising of a reversible and irreversible component~\cite{Nogus1999,Jenkins2020} where the former follows the FM spins while the latter is strongly coupled to the AFM bulk. The different behaviors of these two components are also observed during the field cycling as will be discussed later. In the AFM layer, the spin texture setting propagates from the topmost layer down to the bottom layer with the polarity of the texture alternating between each consecutive monolayer, as was shown in Figure \ref{fig:stack}. This is in a good agreement with the typical tetrahedral 3Q spin state of the AFM ordering of $\gamma$-IrMn$_3$ \cite{Sakuma2003} and [001] crystal ordering. 
To further confirm the alternating magnetization in the antiferromagnet we calculate the partial spin-spin correlation function between one layer of the ferromagnet and each layer of the antiferromagnet, shown in Figure~\ref{fig:ssc}. The spin-spin correlation function is given by
\begin{equation}
    C(\Delta r_z) = \frac{1}{N}\sum_{i,j} s^z_i(r_z) s^z_j(r_z+\Delta r_z)
\end{equation}
where $i$ are atoms in the center ferromagnetic layer (containing the ferromagnetic spin texture), $j$ are atoms in the antiferromagnet separated by a position vector $r_z$ with $z$ components only (we do not consider pairs of atoms separated laterally) and $N$ is the total number of pairs considered. This gives the correlation between the ferromagnet and successive layers of the antiferromagnet, with a value of 1 being full correlation and -1 being fully anti-correlated. In the case of the spin texture the ordering of the interfacial moments of the AFM is highly correlated with the ferromagnetic order, leading to a net and alternating correlation between the FM and each layer of the AFM in the system. In the absence of FM-AFM coupling there is no net correlation and in the case of no spin texture the correlation is similar. The interfacial layer feels a strong exchange field from the ferromagnet leading to a large positive correlation, while the next two subsurface layer suffer frustration due to the intermixing with the ferromagnetic atoms. The bulk correlation is approximately constant but with a small decrease in the strength of correlation towards the bottom of the antiferromagnetic layer due to the loss of surface exchange bonds and a local decrease in antiferromagnetic surface order, in agreement with expectations~\cite{Jenkins2018, Jabakhanji2023, Ga_2023}.

In the bulk of the AFM (IrMn 13 in Figure \ref{fig:spintext}) the noise is reduced, as there is no need to accommodate intermixing and the resulting non-IrMn AFM bonds, but some noise remains from the local nature of the anisotropy of $\gamma$-IrMn$_3$ ~\cite{Jenkins2019}. The behavior of the $s_z$ component follows that of the FM layer, with the only exception being that the domain walls are narrower as they follow the AFM grain boundaries. The penetration depth of the setting, down to the last layer IrMn 34
\begin{figure}[!ht]
\includegraphics[width=0.49\textwidth, trim=0 0 0 0]{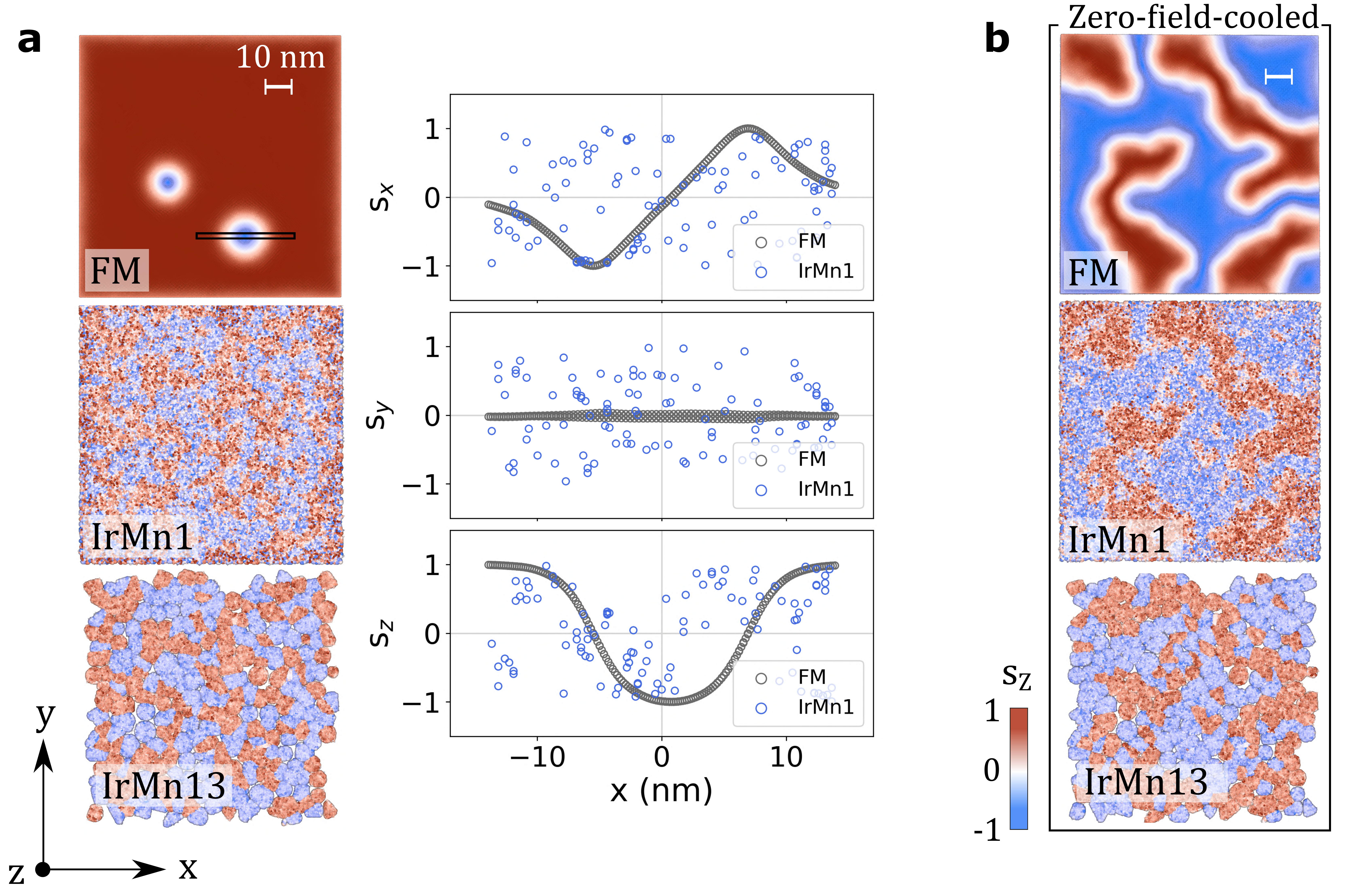}
\caption{(Color Online) a) No imprinting is observed for a simulation system without exchange interaction across the FM/AFM interface. b) Maze domains can be imprinted from the FM layer into the AFM layer under zero-field cooling conditions.}
\label{fig:maze}
\end{figure}
which is 5 nm away from the interface, is in good agreement with experimental reports of vortices in CoO and NiO \cite{Wu2011} and exchange springs in $\gamma$-IrMn$_3$ \cite{Reichlova2016}, which could propagate over the whole AFM thickness. As a control simulation, in Figure \ref{fig:maze}a we show that for a system without FM-AFM interfacial exchange coupling there is no correspondence between $s_i$ of the FM and AFM layers and instead the setting of the AFM grains is random.

\section{Field-stability of the set AF grains}
\begin{figure}[!ht]
\includegraphics[width=0.43\textwidth, trim=0 0 0 0]{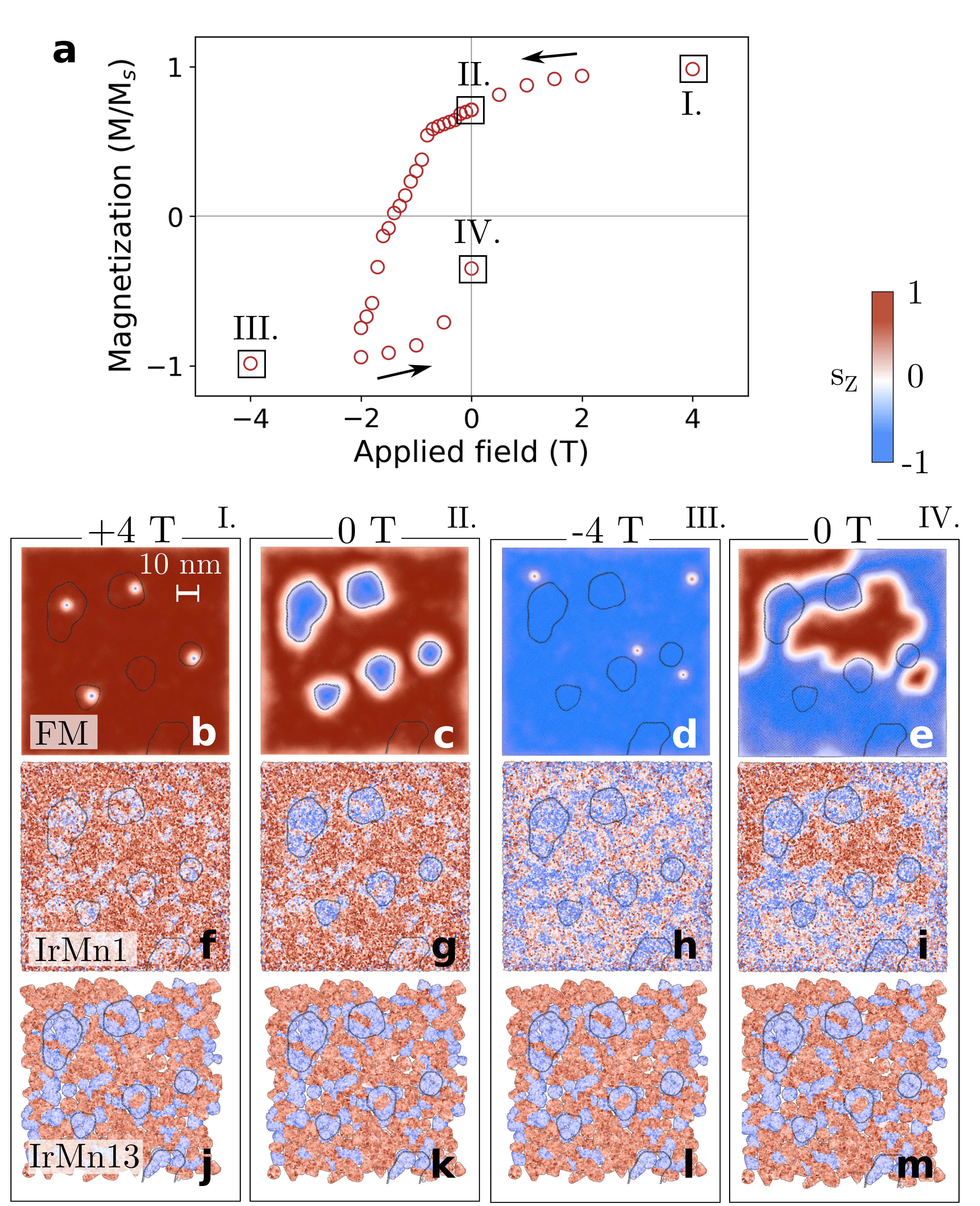}
\caption{(Color Online) (a) Field-dependence of the FM magnetization (M/Ms) at 0 K and (b-m) snapshots of the evolution of the spin structures during the field-sweep, in the FM (b-e), AFM at the interface (f-i, IrMn 1) and AFM in the core (j-m, IrMn 13). The snapshots are taken at positive saturation (I.), forward remanence (II.), negative saturation (III.) and backward remanence (IV.). The contours of the FM Sks in \ref{fig:imprinting}n) are superimposed to all images, as visual guides.}
\label{fig:hysteresis}
\end{figure}

Next, we study the effect of cycling the external field (oriented along the film normal) on the spin textures throughout the simulation stack. As expected, the hysteresis loop (Figure \ref{fig:hysteresis}a) shows a shift of $\sim -0.7$ T towards negative fields due to the exchange bias from the uncompensated, irreversible interfacial Mn spins \cite{Jenkins2020}. The spin texture evolution, on the other hand, strongly depends on the layer. The FM layer follows the field, nearly saturating at strong fields (Figure \ref{fig:hysteresis}b,d) and gradually reversing at intermediate fields by the expansion of Sk bubbles. The AFM bulk, on the other hand, is unchanged throughout the field cycle as seen in Figure \ref{fig:hysteresis}j-m. This can also be seen in the fact that at remanence, the FM layer (Figure \ref{fig:hysteresis}c,e) adopts a configuration that conforms with that of the AFM layer. The AFM interface comprises a reversible component that follows the FM and an irreversible component that abides by the AFM bulk, and therefore shows a behavior that is a mixture of the two (Figure \ref{fig:hysteresis}f-l). This shows that the local spin textures that have been set remain unaltered in the AFM bulk and partially at the interface even after the original Sks in the FM layer are annihilated. This means that we can transform a system with Sks in the FM layer and no spin textures in the AFM (Figure \ref{fig:imprinting}a-c) to a system with no Sk in the FM layer
but pre-defined spin textures in the AFM layer (Figure \ref{fig:hysteresis}d-l) using consecutive thermal and field cycling processes. This opens up perspectives for further studies of spintronic properties of isolated, localized spin textures in an AFM.
%\vspace*{-5mm}
\vspace{2mm}
\section{Discussion}
Here we open the discussion on future study of topologically protected states in the real space of compensated magnets as they hold promise of favorable and technologically relevant properties \cite{Jungwirth2016,Baltz2018,Smejkal2018}. These spin textures cannot be continuously transformed into a topologically trivial state and therefore are remarkably stable. Moreover, they impact the physical properties of the system as the winding of the spins allows the electrons to acquire a non-zero geometric phase \cite{Xiao2010,Gbel2018} when interacting with the local spin structure. While our work demonstrates the localized setting of non-topological spin textures in AFM grains, the technique we have presented could be applied and optimized for continuous films of AFMs or compensated magnets with spin-split band structure \cite{Chen2014,Smejkal2022}. The latter group is interesting due to their novel transport phenomena such as spontaneous anomalous Hall effect (AHE) \cite{Nakatsuji2015,Feng2022,Reichlova} and spin current generation \cite{elezn2017,GonzalezHernandez2021,Bose2022} allowed by the novel band topologies. A particular topological state of interest would be Sks with compensated magnetic ordering - due to the anti-parallel interatomic exchange interactions they have not only zero net magnetization but also zero topological charge, which ensure robustness against external fields and vanishing skyrmion \cite{Zhang2016afm,Barker2016} and topological Hall effects \cite{Gobel2017}. Other predicted effects of AFM Sks on the transport properties include the non-vanishing topological spin Hall effect \cite{Gobel2017,Akosa2018} and a longitudinal Sk velocity exceeding that of the FM Sks \cite{Zhang2016afm,Barker2016}. These properties make Sks with compensated magnetic ordering not only fascinating subjects for studies on topology but also competitive information carrier candidates for ultra-dense, ultra-fast, low-power spintronic devices.

\section{Conclusion}

In conclusion, we have theoretically demonstrated the setting of localized spin textures in antiferromagnetic $\gamma$-IrMn$_3$ by using predefined spin textures in an adjacent exchange-coupled ferromagnet as a template. This setting of non-topological spin textures was realized by a thermal cycling procedure and it was shown to extend beyond the interface through the entire thickness of the antiferromagnet (here 5 nm). The set AFM textures showed remarkable stability against field perturbations and the setting efficiency as well as the morphologies of the set AFM textures were shown to depend on various material parameters. This work offers a solution for overcoming the challenge of nucleating localized real-space spin textures in compensated magnets with zero net magnetization, promoting the extension of local spin texture studies beyond ferromagnets. 

\newpage
\section*{Data access statement}
The raw data for this paper were generated at ARCHER2 . Derived data supporting the findings of this study are available from the corresponding authors on request.
\begin{acknowledgments}
This study was partially supported by the France-UK Alliance Hubert Curien program (PHC) (Grant No. 46298XC) and the UK EPSRC program (Grant No. EP/V007211/1). This work used the ARCHER2 UK National Supercomputing Service (https://www.archer2.ac.uk). 
\end{acknowledgments}
%\appendix

%\clearpage
%\bibliography{/Users/rfle500/Documents/Work/Papers/Bibliography/library}
%\bibliography{library,local}
\bibliography{refs.bib}
%---------------------------------------------------------------------------%

%\begin{thebibliography}{11}%
%\end{thebibliography}%

\end{document}